\documentstyle[aps,prd,
amsmath,epsfig,rotating]{revtex}

\begin{document}

\draft

\widetext

\title{Analysis of a three flavor neutrino oscillation fit to recent
Super-Kamiokande data}
\author{Christoph Meier\footnote{Electronic address:
{\tt meier@theophys.kth.se}} and Tommy Ohlsson\footnote{Electronic address:
{\tt tommy@theophys.kth.se}}}
\address{Division of Mathematical Physics, Theoretical Physics, Department of
Physics, Royal Institute of Technology,\\ SE-100 44 Stockholm, Sweden}
\date{\today}

\maketitle

\begin{abstract}
We have analyzed the most recent available Super-Kamiokande data in a
three flavor neutrino oscillation model. We have here neglected possible
matter effects and performed a fit to atmospheric and
solar Super-Kamiokande data. We have investigated a large parameter range,
where the mixing angles were restricted to $0 \leq \theta_i \leq
\pi/2$, $ i=1,2,3$, and the mass squared differences were taken
to be in the intervals $10^{-11}\;{\rm eV}^2 \leq
\Delta m^ 2 \leq 10^{-2}\;{\rm eV}^2$ and $10^{-4} \;{\rm eV}^2 \leq
\Delta M^2 \leq 10 \;{\rm eV}^2$. This yielded a best solution characterized by
the parameter values
$\theta_1 \simeq 45^\circ$, $\theta_2 \simeq 10^\circ$, $\theta_3
\simeq 45^\circ$, $\Delta m^2 \simeq 4.4 \times 10^{-10}\; {\rm eV}^2$, and
$\Delta M^2 \simeq 1.01 \times 10^{-3}\;{\rm eV}^2$, which shows that the
analyzed experimental data speak in favor of a bimaximal mixing
scenario with one of the mass squared differences in the ``just-so'' domain
and the other one in the range capable of providing a solution to the
atmospheric neutrino problem. 
\end{abstract}


\section{Introduction}
\label{kamiointro}

Most of the analyses concerning neutrino oscillations have so far been
done within the framework of the theory of two flavor neutrino
oscillation. In these scenarios, the observed deficit of solar
electron neutrinos is usually explained by means of $\nu_e
\leftrightarrow \nu_\mu$ oscillations, whereas the lack of atmospheric
muon neutrinos is interpreted as a consequence of $\nu_\mu
\leftrightarrow \nu_\tau$ oscillations. The main advantage of the two
flavor oscillation scenario is that the oscillation probability
depends on only two parameters, the mass squared difference $\Delta
m^2 $ and the mixing angle $\theta$.

Recently, it was pointed out by several authors, that in
some cases the interpretation of the experimental data in terms of two flavor
oscillations might yield misleading and sometimes even wrong results
(see {\it e.g.} Ref.\cite{Ohlsson}) and that a three flavor oscillation
description is to be favored. Unfortunately, the oscillation
probabilities depend in this case on five parameters, two mass squared
differences and three mixing angles, which makes it quite hard to deal
with the dependence of the probability functions on these
parameters. 

The most popular three flavor neutrino oscillation scenario is the
so called bimaximal mixing scenario \cite{Barger}, where two of the
mixing angles 
are maximal, {\it i.e.}, they have values of about $45^\circ$, whereas
the remaining one is restricted by the CHOOZ experiment to be rather
small at least for certain parameter ranges \cite{Bilenky}. The three
flavor oscillation scenario can then be shown to decouple into two
distinct scenarios involving two flavors, in which case it is of
course much easier to 
perform a fit to the experimental data. Many authors \cite{Petcov} use
the CHOOZ result to simplify the oscillation probability formulas for three
flavors and then perform a fit of the two decoupled two flavor
neutrino oscillation scenarios to the experimental data.

In this paper, we will present a numerical fit within
a three flavor neutrino mixing scenario to the zenith angle distribution of 850
day atmospheric neutrino Super-Kamiokande data and 708 day solar
neutrino Super-Kamiokande data. However, we will not make any
assumptions about the parameters on which the oscillation probability
formulas depend, {\it i.e.}, we fit the scenario in the most general case and
for a large parameter range to the experimental data.

The paper is organized as follows. In
Sec.~\ref{kamiotheory}, we summarize the basic features of three
flavor neutrino oscillation theory. In
Sec.~\ref{kamioexp}, we present the choice of experimental data and in
Sec.~\ref{kamiomini}, we discuss the minimization procedure. In
Sec.~\ref{kamioresults}, we give the obtained solutions of the
minimization problem, which are interpreted in
Sec.~\ref{kamiointerpretation}. The fits corresponding to the
solutions are discussed and compared in Sec.~\ref{kamiofits} and then
these solutions are tested for stability in
Sec.~\ref{kamiostability}. Finally, Sec.~\ref{kamioconclusions} 
contains a summary and our conclusions.

\section{Theory of Three Flavor Neutrino Oscillations}
\label{kamiotheory}

In three flavor neutrino oscillation theory one assumes the neutrino
states with definite flavor $\vert \nu_\alpha \rangle$, $\alpha =
e,\mu,\tau$, to be linear superpositions of states with definite mass
$\vert \nu_i \rangle$, $i=1,2,3$,
\begin{equation}
\vert \nu_\alpha \rangle = \sum_{i=1}^3 U_{\alpha i}^* \vert \nu_i
\rangle, \quad \alpha=e,\mu,\tau.
\end{equation}
The unitary mixing matrix $U$ is called Cabibbo--Kobayashi--Maskawa
(CKM) matrix and can generally be parameterized by three mixing angles
$\theta_i$, where $i=1,2,3$, and three CP-violating phases. One of the
latter were recently 
claimed to be observable by next generation neutrino detectors
\cite{Lindner} and could lead to physically very interesting
consequences. The other two can be shown to cause no physical
effects. However, for this analysis these phases are too small
to be important and we will neglect them in what follows. This implies
$U_{\alpha i}=U_{\alpha i}^*$ and therefore we can write the CKM matrix
in its (real) standard parameterization \cite{CKM} as
\begin{equation}
U=\left(
\begin{matrix}
c_2 c_3 & s_3 c_2 & s_2 \\ -s_3 c_1-s_1 s_2 c_3& c_1 c_3-s_1 s_2 s_3 &
s_1 c_2\\ s_1 s_3-s_2 c_1 c_3 & -s_1 c_3-s_2 s_3 c_1 & c_1 c_2
\end{matrix}
\right),
\end{equation}
where $s_i \equiv \sin\theta_i$ and $c_i \equiv \cos\theta_i$.
The probability for detection of a neutrino of flavor $\beta$ in a
beam of neutrinos consisting exclusively of flavor $\alpha$ at the
source of the beam is for three flavors given by
\begin{equation}
P_{\alpha \beta}= \delta_{\alpha \beta}-4\mathop{\sum_{i=1}^3
\sum_{j=1}^ 3}_{i<j} U_{\alpha i} U_{\beta i} U_{\alpha j} U_{\beta j}
\sin^ 2\left(\frac{\Delta m^2_{ij} L}{4E}\right), \quad
\alpha,\beta=e, \mu, \tau,
\end{equation}
where $\delta_{\alpha \beta}$ is Kronecker's delta, $L$ is the
distance from the source to the detector, $E$ is the neutrino energy, and
$\Delta m^2_{i j}$ is the difference between the squares of the masses
corresponding to the mass eigenstates $\vert \nu_i \rangle$ and $\vert \nu_j
\rangle$. Since
\begin{equation}
\Delta m^2_{21} + \Delta m^2_{32} + \Delta m^2_{13} = 0
\end{equation}
the mass squared differences are not linearly independent, and it is
convenient to choose
\begin{eqnarray}
\Delta m^2 &\equiv& \Delta m^2_{21},\\ \Delta M^2 &\equiv& \Delta m^2_{32},
\end{eqnarray}
from which follows that
\begin{equation}
\Delta m^2_{31}=-\Delta m^2_{13}=\Delta m^2 + \Delta M^2.
\end{equation}

So far we have considered the neutrino states to be plane waves,
{\it i.e.}, to have a definite momentum. Since the neutrino state is
neither produced nor detected with a definite momentum or propagation
length, one has to average over $L/E$ as well as other uncertainties
in production and detection. We will here follow closely Ref. \cite{Ohlsson}
and assume that these uncertainties to be described by a Gaussian average
\begin{equation}
\langle P _{\alpha \beta} \rangle = \int_{-\infty}^{\infty} P_{\alpha
\beta}(x) f(x) dx,
\end{equation}
where
\begin{equation}
f(x)=\frac{1}{\gamma\sqrt{2\pi}}\exp
\left\{-\frac{(x-l)^2}{2\gamma^2}\right\}
\end{equation}
and $x \equiv L/E$. Inserting formula (3) for $P_{\alpha \beta}$ yields
\begin{equation}
\langle P_{\alpha \beta} \rangle = \delta_{\alpha \beta} - 2
\mathop{\sum_{i=1}^3 \sum_{j=1}^3}_{i<j} U_{\alpha i} U_{\beta i}
U_{\alpha j} U_{\beta j} \left[ 1 - \cos(2l\Delta
m^2_{ij})\exp\{-2\gamma^2(\Delta m^2_{ij})^2\} \right],
\label{probability2}
\end{equation}
where $l$ is given by
\begin{equation}
l \simeq 1.27 \left \langle \frac{L}{E} \right \rangle \simeq
1.27\frac{\langle L \rangle }{\langle E \rangle}
\end{equation}
and therefore related to the sensitivity of the experiment. $\langle L
\rangle$ should here be inserted in meters, whereas $\langle E
\rangle$ should be inserted in MeV. The parameter $\gamma$ is the so
called damping factor and in accordance with Ref. \cite{Ohlsson} we will
choose
\begin{equation}
\gamma=\frac{L}{E}\left(\frac{\Delta L}{L}+\frac{\Delta E}{E}\right),
\end{equation}
where $\Delta E$ and $\Delta L$ are the uncertainties in neutrino energy and
propagation length, respectively.

For large values of $\gamma$
$$
\exp\{-2\gamma^2(\Delta m^2_{ij})^2\} \rightarrow 0
$$
and the oscillation term vanishes. The transition probability then
becomes a constant dependent on the values of the mixing angles; the
oscillation is said to become ``washed out''.  To proceed, one now has
to determine the values of the parameters $L,E,\Delta L,$ and $\Delta
E$ for the experiments under consideration.

\section{Choice of Experimental Data}
\label{kamioexp}

We are going to consider three types of Super-Kamiokande data, multi-GeV
$e$-like atmospheric neutrino data, multi-GeV $\mu$-like atmospheric
neutrino data, and solar neutrino data.

\subsection{Solar neutrino data}

The probabilities were taken from Ref. \cite{Berezinsky} for 14 data points
in an energy range from 6 MeV to 12 MeV. The probability is almost
constant for all 14 data points.  The path length for solar 
neutrinos is given by the distance from the Sun to Earth ($L \simeq
1.44 \times 10^{11}$ m) and the uncertainty of the path length is here
assumed to be negligible compared to $L$, {\it i.e.},
\begin{equation}
\frac{\Delta L}{L} \simeq 0.
\end{equation}
The damping factor then becomes
\begin{equation}
\gamma \simeq \frac{L}{E}\frac{\Delta E}{E},
\end{equation} 
where $\Delta E = 0.5\; {\rm MeV}$ is the energy resolution of the
experiment.

\subsection{Atmospheric neutrino data}

The atmospheric neutrino data are divided into atmospheric multi-GeV
$e$-like events and atmospheric multi-GeV $\mu$-like events. In both
cases, the flux measurements are performed for five bins, {\it i.e.}, for
five different values
$$
\cos\theta= -0.8,-0.4,0,0.4,0.8
$$
of the zenith angle $\theta$, which is defined as the angle between
the direction of the measurement and the axis through experiment and the
middle of the Earth. Therefore, $\cos\theta =-1$ corresponds to
straight upward going neutrinos, whereas for $\cos\theta =1$ the
neutrinos are produced in the atmosphere right above the detector.

The experimental data used are 850 day Super-Kamiokande data and Monte
Carlo no-oscillation predictions for 40 years generated lifetime for
the five bins. The corresponding errors were taken from Ref. \cite{Scholberg}.

The results of the Super-Kamiokande
experiment are usually presented by means of the ``ratio of the
ratios''
\begin{equation}
{\cal R}=\frac{\Phi_\mu}{\Phi_e}(data)/\frac{\Phi_\mu}{\Phi_e}(MC),
\end{equation}
where $\Phi_\alpha$ is the flux of neutrino flavor $\alpha$.

In Summer 1998, the Super-Kamiokande Collaboration published
experimental values for this 
``ratio of the ratios'' of ${\cal R}=0.63 \pm 0.03(stat) \pm 0.05(sys)$ for
sub-GeV events and ${\cal R}=0.65 \pm 0.03(stat) \pm 0.08 (sys)$ for multi-GeV
events \cite{Superk}, which was the first direct observation of
neutrino oscillations.

However, to be able to fit both, $e$-like and
$\mu$-like data to the measurements, we will use here the directly
measured disappearance rates
\begin{equation}
R_e=\frac{\Phi_e(data)}{\Phi_e(MC)}
\end{equation}
for $e$-like events and
\begin{equation}
R_\mu=\frac{\Phi_\mu(data)}{\Phi_\mu(MC)}
\end{equation}
for $\mu$-like events. The main disadvantage of this is the
uncertainty in the expected neutrino flux, which was claimed to be up
to 20$\%$ \cite{Takaaki}, due to the unknown flux of cosmic particles
hitting the atmosphere. To handle this one can introduce an overall
flux uncertainty factor $ \eta $ with values between 0.8 and 1.2. The
actual neutrino flux for the neutrino flavor $\alpha$ is then given by
$\tilde{\Phi}_\alpha = \eta \Phi_\alpha$. Thus, in absence of neutrino
oscillations, the ratio between the theoretical and the measured fluxes would
just be $R_\alpha = \eta$, whereas otherwise one obtains
\begin{equation}
R_\alpha=\frac{\tilde{\Phi}_e P_{e\alpha}+\tilde{\Phi}_\mu
P_{\mu\alpha}+\tilde{\Phi}_\tau P_{\tau\alpha}}{\Phi_\alpha},
\quad \alpha=e,\mu,\tau,
\end{equation}
where $P_{\alpha\beta}$ is the transition probability from $\alpha$ to
$\beta$, see Sec.~\ref{kamiotheory}. In what follows, we are going to
neglect the $\tau$ contribution due to too small production cross
sections. The uncertainty in the flux of cosmic rays is then an
overall factor common to the ratios $R_\alpha$, where $\alpha = e,\mu$, of
the both
remaining neutrino flavors. Denoting $R \equiv \Phi_\mu/\Phi_e$, one obtains
\begin{equation}
R_e=\eta(P_{ee}+R P_{\mu\mu})
\label{disappearencerate_e}
\end{equation}
for the electron neutrino ratio and
\begin{equation}
R_\mu=\eta \left(P_{ee}+\frac{1}{R}P_{\mu\mu}\right)
\label{disappearencerate_mu}
\end{equation}
for the muon neutrino ratio. The value of $R$ is theoretically well
determined to be about $R \simeq 3$ at $E \simeq 10$ GeV \cite{Gaisser}.

The energy of the measured multi-GeV neutrinos is assumed to be $E=10$
GeV with negligible uncertainty compared to $E$ and we set
$\Delta E/E \simeq 0$.
The path length for a neutrino that was detected in a bin
corresponding to the zenith angle $\theta$ is given by
\begin{equation}
L(\theta)=\sqrt{r^2 \cos^2\theta +2 r d + d^2} - r \cos\theta,
\end{equation}
as is easily obtained from geometrical considerations. Here $r$ is the
radius of the Earth and $d$ is the typical altitude of the neutrino
production point in the atmosphere, which we assumed to be $d = 10$
km. The uncertainty in path length is mainly
determined by $\Delta\cos\theta = 0.2$ and therefore
\begin{equation}
\Delta L = \left\vert\frac{\partial L(\theta)}{\partial
\cos\theta}\right\vert \Delta\cos\theta = \frac{r L}{L+r \cos\theta}\;
\Delta\cos\theta 
\end{equation}
The damping factor $\gamma$ is thus a function of $\theta$
\begin{equation}
\gamma \simeq \frac{\Delta L}{E}.
\end{equation}

\section{The Minimization Procedure} 
\label{kamiomini}

According to the last section, we have 24 data points, which we can use
for the fit to the parameters of the theory: the five bins for $e$-like
events, the five bins for $\mu$-like events, and the 14 data points for
the solar neutrinos.

The function used to obtain the parameters is given by
\begin{eqnarray}
\chi(\theta_1,\theta_2,\theta_3,\Delta m^2,\Delta M^2)&=&\frac{1}{{\rm
w}_{tot}}\bigg\{\sum_{i=1}^5 {\rm w}_{e,i}
\frac{1}{R^2_{e,i}(exp)} \left[R_{e,i}-R_{e,i}(exp)\right]^2
+\sum_{i=1}^5 {\rm
w}_{\mu,i}\frac{1}{R^2_{\mu,i}(exp)}
\left[R_{\mu,i}-R_{\mu,i}(exp)\right]^2\nonumber\\ 
&+&\sum_{i=1}^{14}{\rm
w}_{sun,i}\frac{1}{P^2_{sun,i}(exp)}
\left[P_{sun,i}-P_{sun,i}(exp)\right]^2\bigg\},
\label{minfunction}
\end{eqnarray} 
where the first two sums run over the five bins and the third one over
the 14 data points for the solar neutrinos. The weights have been
taken as
\begin{eqnarray}
{\rm w}_{e,i} &=& R_{e,i}(exp)/\Delta R_{e,i}(exp), 
\label{eweight}\\
{\rm w}_{\mu,i} &=& R_{\mu,i}(exp)/\Delta R_{\mu,i}(exp),
\label{muweight}\\
{\rm w}_{sun,i} &=& P_{sun,i}(exp)/\Delta P_{sun,i}(exp), 
\label{sunweight}\\
{\rm w}_{tot} &=&
\sum_{i=1}^5 {\rm w}_{e,i}+\sum_{i=1}^5 {\rm
w}_{\mu,i}+\sum_{i=1}^{14}{\rm w}_{sun,i}.
\label{weight}
\end{eqnarray}
The mixing angles were constrained to the interval $[0,\pi/2]$,
whereas for the mass squared differences we assumed one of them to be in the
range $10^{-11} \; {\rm eV}^2 \leq \Delta m^2 \leq 10^{-2} \; {\rm eV}^2$
and the other one in $10^{-4} \; {\rm eV}^2 \leq \Delta M \leq 10 \;
{\rm eV}^2$. Since these large parameter ranges yield no numerically
stable solution, we minimized function (\ref{minfunction}) for 45
parameter ranges corresponding to all combinations of values for the
mass squared differences between
\begin{eqnarray}
10^{-11} \; {\rm eV}^2 &\leq& \Delta m^2 \leq 10^{-10} \; {\rm eV}^2,
\nonumber\\ 
10^{-10} \; {\rm eV}^2 &\leq& \Delta m^2 \leq 10^{-9} \; {\rm
eV}^2,\nonumber\\ 
&\vdots& \hspace{1cm}\vdots\nonumber\\ 10^{-3} \;
{\rm eV}^2 &\leq &\Delta m^2 \leq 10^{-2} \; {\rm eV}^2 \nonumber
\end{eqnarray}
for the small mass squared difference and
\begin{eqnarray}
10^{-4} \; {\rm eV}^2 &\leq& \Delta M^2 \leq 10^{-3} \; {\rm
eV}^2,\nonumber\\ 10^{-3} \; {\rm eV}^2 &\leq& \Delta M^2 \leq 10^{-2}
\; {\rm eV}^2, \nonumber\\ &\vdots& \hspace{1cm}\vdots\nonumber\\ 1 \;
{\rm
eV}^2 &\leq& \Delta M^2 \leq 10 \; {\rm eV}^2\nonumber
\end{eqnarray}
for the larger one.

Following Ref. \cite{Ohlsson}, we chose the following
strategy to deal with the flux uncertainty factor $\eta$: During the
search for minima of function~(\ref{minfunction}), we set it equal
to unity, {\it i.e.}, we neglected the flux uncertainty. The solutions
obtained in this way were then tested for their stability with respect
to variations of $\eta$, see Sec.~\ref{kamiostability}.

To minimize the function
$\chi$ in Eq.~(\ref{minfunction}) we generated $N=10^6$ random values
for each of the five parameters in the corresponding range such that we
obtained $N$ values of the minimization function. After that we picked
the parameter set that yielded the smallest value for the function and
repeated this procedure $n=20$ times. Each of the obtained $n$
parameter sets then served as starting values for a deterministic
minimization procedure, using a sequential quadratic programming method
such that one obtains again $n$ parameter sets as an end result.

\section{Solutions of the Minimization Problem}
\label{kamioresults}

The solutions to the minimization problem can best be obtained by
considering tables like Table~\ref{masstable}, where the best point
values for the two mass squared differences for each of the
investigated mass regions are shown. One can see, that for most of the
mass regions at least one of the two mass squared differences was
obtained on one of the region boundaries. In this case, we assumed that
the value of the corresponding mass squared difference converges
towards a better minimum in one of the neighbor regions. Thus, a given
parameter region was only considered to contain a solution to the
minimization problem, if both values of the mass squared differences
were obtained within the boundaries of the region. 
Note, that the values obtained for $\Delta M^2$ are almost
randomly distributed in the last column of Table~\ref{masstable}, {\it
i.e.}, in the range $1\;{\rm eV}^2\leq \Delta M^2 \leq 10 \;{\rm eV}^2$. This
is due to a too large value of the damping factor $\gamma$ in
Eq.~(\ref{probability2}) such that the minimization function becomes
independent of this parameter. Only $\Delta m^2$ and the mixing angles are in
this case fitted to the experimental data.

To keep overview over the various considered parameter regions one can
consider Table~\ref{masstable2}. Here a region, where a value for $\Delta
M^2$ was obtained within the boundaries, is marked with an ``$M$'',
whereas a region with a corresponding value for $\Delta m^2$ is marked
with an ``$m$''. One obtains in this way eight regions, marked
``$M\;m$'' in Table~\ref{masstable2}, where the mass squared
differences are within the region boundaries. Seven of these regions
are contained in the first two rows of Table~\ref{masstable2}, whereas
the eighth one was obtained within the range $10^{-4}\;{\rm eV}^2\leq
\Delta m^2 \leq 10^{-3}\;{\rm eV}^2$, $1\;{\rm eV}^2\leq \Delta
M^2\leq 10\;{\rm eV}^2$. The parameter values obtained from the
minimization procedure with the mass squared differences constrained to
these regions correspond to the possible solutions of the minimization
problem. One notices that the regions marked with
``$M\;m$'' split Table \ref{masstable2} in three distinct areas,
corresponding to three types of solutions, as we will see.

The first of these interesting areas is the one containing the
parameter regions in the upper left corner of Table \ref{masstable2},
{\it i.e.}, the nine 
regions in the range $\; 10^{-11}\;{\rm eV}^2\leq \Delta m^2\leq
10^{-8}\;{\rm eV}^2$, $10^{-4}\;{\rm eV}^2\leq \Delta M^2\leq
10^{-1}\;{\rm eV}^2$. In this range, there are all together five
possible solutions, {\it i.e.}, mass regions, where the mass
squared differences were obtained within the region boundaries. To pick
the best of the possible solutions for this 
type, consider Table \ref{minimizationtable}, where the best point
values for the minimization function (\ref{minfunction}) are
shown. The smallest value of the minimization function was obtained
in the region corresponding to $10^{-10}\;{\rm eV}^2\leq \Delta m^2 \leq
10^{-9}\;{\rm eV}^2$, $10^{-3}\;{\rm eV}^2\leq \Delta M^2\leq
10^{-2}\;{\rm eV}^2$. In all surrounding regions, we obtained
values for the minimization function, which are larger, such that there
seems indeed to be a minimum in the region specified above. Thus, the
corresponding parameter set is a solution of the minimization problem,
and we will denote it by {\it Solution 1} in what follows.
Considering the rest of Table \ref{minimizationtable}, one can see
that this solution corresponds to the smallest value of the
minimization function in the whole investigated parameter range, which
means that it provides the best fit to the experimental data, as will
be seen in Sec. \ref{kamiofits}. Note that the value of the
minimization function in the parameter range containing Solution 1 is
very close to the one obtained in the region $10^{-10}\;{\rm eV}^2\leq \Delta
m^2 \leq 10^{-9}\;{\rm eV}^2,\; 10^{-4}\;{\rm eV}^2\leq \Delta M^2\leq
10^{-3}\;{\rm eV}^2$. Furthermore, the values for the mass 
squared differences obtained within these two regions are rather close
to each other, and one can not really distinguish, if these two
possible solutions correspond to the same minimum or to two distinct
minima very close to each other.

The second interesting area of
Table \ref{masstable2} contains six parameter regions in the range
$10^{-5}\;{\rm eV}^2\leq \Delta m^2 \leq 10^{-3}\;{\rm eV}^2$,
$10^{-1}\;{\rm eV}^2\leq \Delta M^2\leq 10\;{\rm eV}^2$. Here a
second type of solution is obtained in the region $10^{-4}\;{\rm
eV}^2\leq \Delta m^2 \leq 10^{-3}\;{\rm eV}^2$, $1\;{\rm eV}^2\leq
\Delta M^2\leq 10\;{\rm eV}^2$, {\it i.e.}, in the last column, eighth row of
Table \ref{masstable2}. We will denote this solution by {\it Solution
2} in what follows. Considering now Table \ref{minimizationtable}, one
can see that in one of the neighbor regions we obtained a
smaller value for the minimization function. One has therefore only a
``local'' minimum inside the region containing Solution 2, and the
values of the minimization function in the surrounding regions do not
converge that obviously towards this region, as they did in the case of
Solution 1. However, in all the surrounding regions at least one of
the mass squared differences was obtained on one of the region
boundaries, which is why we will consider the region $10^{-4}\;{\rm
eV}^2\leq \Delta m^2\leq 10^{-3}\;{\rm eV}^2$, $1\;{\rm eV}^2\leq
\Delta M^2\leq 10\;{\rm eV}^2$ to contain the second solution of the
minimization problem. Comparing the corresponding value of the
minimization function to the one of Solution 1, one sees that this
second solution provides a less exact fit to the experimental data.

Finally, the third interesting area is situated in the upper right
corner of Table \ref{masstable2}, in the range $10^{-11}\;{\rm eV}^2\leq
\Delta m^2\leq 10^{-9}\;{\rm eV}^2$, $10^{-1}\;{\rm
eV}^2\leq \Delta M^2\leq 10\;{\rm eV}^2$. Here two regions with both mass
squared differences within the region boundaries are
obtained. Consideration of Table \ref{minimizationtable} tells us,
that the smallest value for the minimization function was obtained in
the parameter region $10^{-10}\;{\rm eV}^2\leq \Delta m^2 \leq 10^{-9}\;{\rm
eV}^2$, $1\;{\rm eV}^2\leq \Delta M^2\leq 10\; {\rm
eV}^2$. Thus, we obtain a third type of solution of the minimization
problem, denoted by {\it Solution 3} in what follows. As in the case
of Solution 2, there are neighbor regions with smaller values for the
minimization function, but they both have at least one of the mass
squared differences on one of the region boundaries. Note that the value
of the minimization function is about five times larger for Solution 3
than the one corresponding to Solution 1, and roughly three times
larger than the one corresponding to Solution 2.

\section{Interpretation of the Obtained Solutions}
\label{kamiointerpretation}

The parameter values for the three solutions obtained from the
analysis in the last section are shown in Table \ref{solutions}. Here
the average value obtained from the $n=20$ minimizations are depicted,
together with the corresponding standard deviations as well as the best point
values. Solution 1 actually corresponds to the
most common three flavor neutrino oscillation scenario, with one of
the mass squared differences in the ``just-so'' domain ($ 10^{-11} {\rm
eV}^2 - 10^{-9} {\rm eV}^2$) and the other one in the range of
possible solutions of the atmospheric neutrino puzzle ($ 10^{-4} {\rm
eV}^2 - 10^{-2} {\rm eV}^2$). As
for our solution, in this common scenario the values of the mixing
angles are normally a set with a small value for $\theta_2$, and
values around $45^\circ$ for the other two mixing angles, $\theta_1$ 
and $\theta_3$, {\it i.e.}, one has bimaximal mixing. The three flavor
oscillation scenario can in this case be shown to decouple into two
independent oscillation scenarios involving two flavors
\cite{Bilenky}. The two flavor oscillation scenario with the small
mass squared difference is then commonly used to describe the solar
electron neutrino deficit in terms of $\nu_e$-$\nu_\mu$ oscillations,
whereas the scenario with the larger mass squared difference is assumed
to describe the atmospheric muon neutrino deficit in terms of
$\nu_\mu$-$\nu_\tau$ oscillations. However, this approximation is only
exact in the case $\theta_2 \rightarrow 0$, and the nonzero value of
this angle causes a mixing of these two decoupled two flavor
scenarios. This influences in turn the values obtained for the mass
squared differences, which might explain the deviation of our result
for the large mass parameter $\Delta M^2=1.01\times10^{-3}\;{\rm
eV}^2$ from the Super-Kamiokande result $\Delta M^2=3.5
\times10^{-3}\;{\rm eV}^2$ \cite{Scholberg}, which was obtained in the
framework of a two flavor oscillation scenario. Note that we obtained
this bimaximal solution without putting any restrictions on the
parameters. Furthermore, the value we obtained for the second mixing
angle $\theta_2 \simeq 10^\circ$ implies that $\sin^2\theta_2
\simeq 0.03$, which is in accordance with the upper bound obtained
from the CHOOZ experiment 
$\sin^2\theta_2 \leq 0.05$ \cite{Bilenky}. A disadvantage of
Solution 1 is the rather large standard deviation of the value
obtained for $\Delta m^2$, see Table \ref{solutions}.

Solution 2, on the other hand, is characterized by a value for the
minimization function, which is twice as large as the one
corresponding to Solution 1, but all parameters, except for the large
mass squared difference, were obtained with very small standard
deviations. As discussed earlier, the minimization function is in this
mass range independent of $\Delta M^2$. This means,
that one obtains random numbers for this parameter, which explains the
large standard deviation of $\Delta M^2$. Only the mixing angles and $\Delta
m^2$ are in this case fitted to the experimental data. Here the small
mass squared difference is lying in the range of possible solutions to
the atmospheric neutrino puzzle, whereas the large one is compatible
with the results of the Liquid Scintillator Neutrino Detector (LSND)
experiment \cite{LSND}, which yielded 
a value for the mass squared difference in the eV-range within the
framework of a two flavor oscillation scenario. However, one can see
from Table \ref{solutions}, that the value obtained for the second
mixing angle is maximal, {\it i.e.}, $\theta_2 \simeq 45^\circ$ and,
unlike for the first solution, it is therefore in this case not
possible to reduce the three flavor oscillation scenario to two
oscillation scenarios involving two flavors.

Finally, Solution 3 has the major disadvantage of a rather large value of the
minimization function compared to the other two solutions, as
discussed in the last section. The remarkably large standard deviation of
the large mass squared difference can be explained as in the case of
the Solution 2, but Solution 3 also shows a rather large standard
deviation of the third mixing angle $\theta_3$. This solution is
actually an ``intersection'' of the first two solutions, with one of
the mass squared differences in the ``just-so'' range and the other one
in the range obtained by LSND. All three mixing angles have values
around $\theta_i \simeq 30^\circ$, where $i=1,2,3$, which means
that it is, like in the case of Solution 2, not possible to decouple
the three flavor oscillation scenario into a pair of two flavor
oscillation scenarios.

\section{Discussion of the Obtained Fits}
\label{kamiofits}

\begin{figure}
\begin{center}
\epsfig{figure=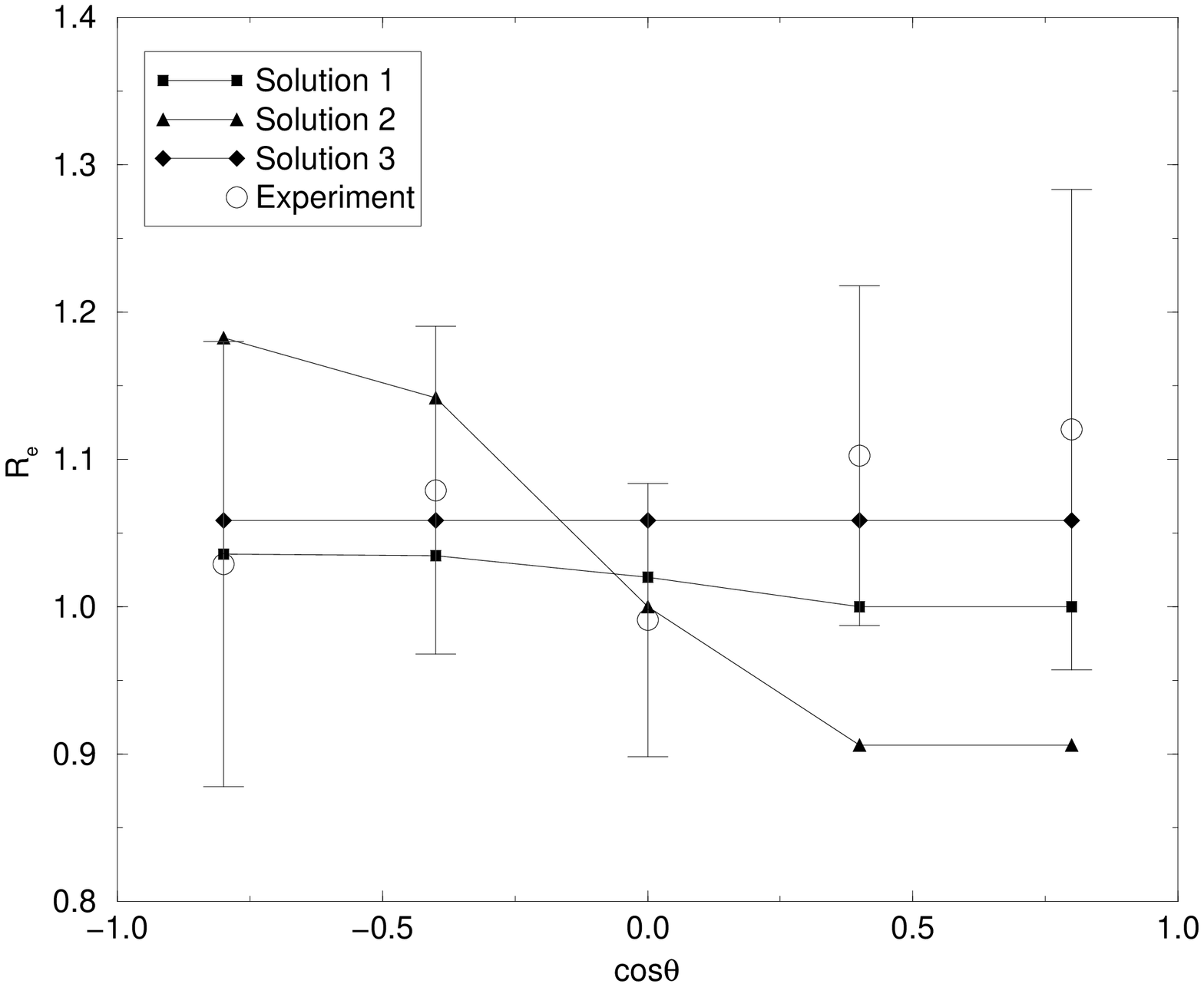,angle=-90,width=7cm}
\caption{The dependence of the ratio $R_e$ on $\cos \theta$.}
\label{diagramatmos_e}
\end{center}
\end{figure}

\begin{figure}
\begin{center}
\epsfig{figure=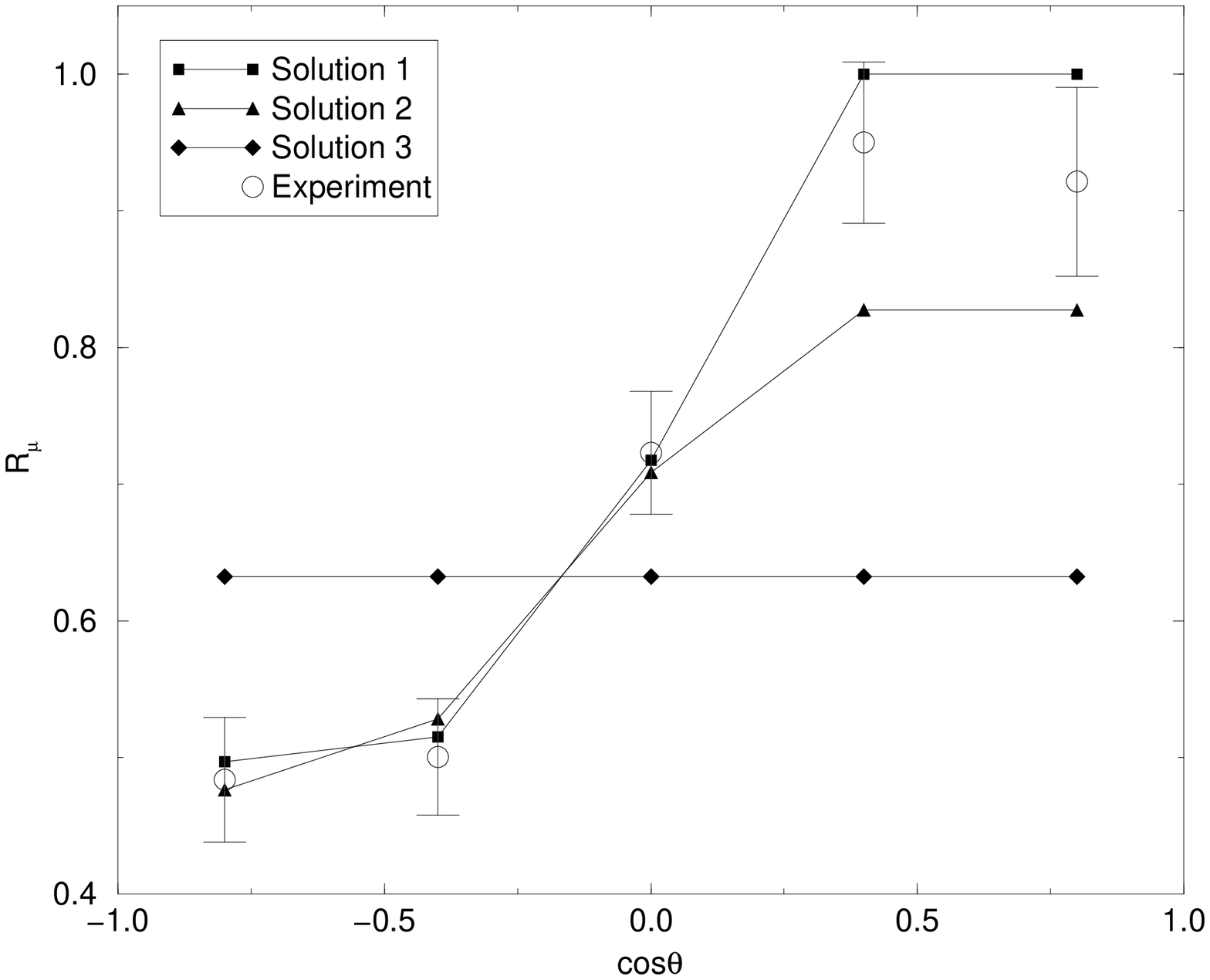,angle=-90,width=7cm}
\caption{The dependence of the ratio $R_\mu$ on $\cos \theta$.}
\label{diagramatmos_mu}
\end{center}
\end{figure}

\begin{figure}
\begin{center}
\epsfig{figure=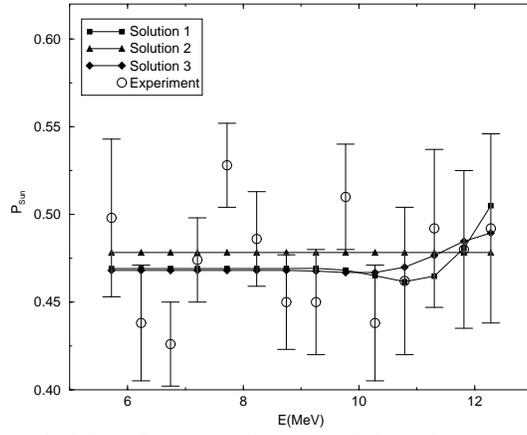,angle=-90,width=7cm}
\caption{The probability $P_{sun}$ as a function of the solar neutrino
energy $E$.}
\label{diagramsolar}
\end{center}
\end{figure}

Figures \ref{diagramatmos_e} - \ref{diagramsolar} show the results of
the fit. The best point probabilities obtained from the three
solutions are depicted as well as compared to the corresponding
experimental data including error bars.

Let us first consider the
results of the fit to the atmospheric electron neutrino data, shown in
Fig. \ref{diagramatmos_e}. Here Solution 1 and Solution 3 provide
the best fits to the experimental data. Solution 2 shows larger
deviations, especially for the two bins with the smallest zenith
angles, {\it i.e.}, the largest values of $\cos\theta$. The neutrinos
measured in these two bins come from right above the detector, which
means that the background of cosmic particles is not shielded by the
Earth. The experimental data corresponding to these zenith
angles have accordingly the largest experimental deviations, and from
Eqs.~(\ref{eweight}) - (\ref{sunweight}) one can see, that these bins
therefore correspond to the smallest weights. This explains why the
fit is less exact for these zenith angles. The fact that one obtains a
zenith angle independent fit for Solution 3 can be understood from
consideration of the oscillation lengths 
\begin{equation}
L_{ij}=\frac{4\pi E}{\vert \Delta m_{ij}^2 \vert}, \qquad i,j=1,2,3,
\quad i \neq j,
\end{equation}
where $E$ is again the neutrino energy and the $\Delta m_{ij}^2$ are the
mass squared differences. This solution has one of the mass squared
differences in the ``just-so'' 
range and the other one in the eV-range. The oscillation length
corresponding to the small mass squared difference is then far too long
to make it possible for the experiment to see any variations in the
oscillation probability, whereas in the case of the large mass squared
difference the oscillations wash out before the neutrinos reach the
detector, due to a too large value of the damping factor $\gamma$ in
Eq.~(\ref{probability2}).

Considering next the fit to the atmospheric muon neutrino data, shown
in Fig. \ref{diagramatmos_mu},
one clearly sees that here Solution 1 provides the best fit. Solution
2 has again larger deviations for the two bins with the smallest
zenith angle, which as before can be explained by the fact that these bins
have the largest experimental errors. Solution 3 provides again a
constant fit to the experimental data, for the same reasons as the
ones discussed above. It appears that this solution is definitely not
capable of explaining the zenith angle dependence of the atmospheric
muon neutrino flux measured by the Super-Kamiokande detector, and
therefore this solution can be ruled out.

Finally, all three solutions provide rather good fits to the solar
neutrino data, see 
Fig. \ref{diagramsolar}. Solution 1 and Solution 3 are seen to yield
a constant oscillation probability up to an energy of about $9\;{\rm
MeV}$. Above that energy the oscillation length becomes of the same order
of magnitude as the Sun-Earth distance such that the oscillation
probability obtained from the two solutions starts to vary with
energy.  Solution 2 yields a constant fit to the solar neutrino data, 
which can be understood, having in mind that this solution is
characterized by one mass parameter in the range $10^{-4}\;{\rm
eV}^2\leq \Delta m^2 \leq 10^{-3}\;{\rm eV}^2$ and the other one in
the eV-range. In both cases, the corresponding oscillation lengths are
much shorter than the Sun-Earth distance, and all oscillatory effects
wash out before the neutrinos reach the Earth.

To summarize, only the Solution 1 provides a good fit to all three types of
experimental data, whereas Solution 3 can be
definitely ruled out by the zenith angle dependence of the atmospheric
muon neutrino data measured by Super-Kamiokande. It remains now to
test the stability of the obtained solutions with respect to variation of
the flux uncertainty factor introduced in Sec. \ref{kamiomini}.

\section{Stability of the Solutions with Respect to the Flux Uncertainty}
\label{kamiostability}

As mentioned before, we performed the minimization procedure setting
the flux uncertainty factor $\eta$ introduced in
Eq.~(\ref{disappearencerate_e}) and Eq.~(\ref{disappearencerate_mu})
equal to one. But this factor can vary between $0.8 \leq \eta \leq
1.2$ and therefore one has to test the solutions obtained in the last
section for stability with respect to variations of this flux
uncertainty factor.

We saw in the last section, that the solution we
denoted by Solution 3 in Table \ref{solutions} can be ruled out by the
zenith angle dependence of the atmospheric muon neutrino flux measured
by Super-Kamiokande. We will therefore only test the stability of
Solution 1 and Solution 2.  To do this, we applied basically the same
minimization procedure as the one described in Sec.
\ref{kamiomini}, but this time for nine equidistant values of the flux
uncertainty factor between $\eta =0.8$ and $\eta = 1.2$. Again, the
angles were restricted to the intervals $0 \leq \theta_i \leq \pi/2$,
where $i=1,2,3$.
  
In the case of Solution 1, the small mass squared
difference was
constrained to $10^{-10}\;{\rm eV}^2\leq \Delta m^2 \leq 10^{-9}\;{\rm
eV}^2$. To choose an interval for the large mass squared difference,
one has to recall that in one of the neighboring regions of Solution 1,
we obtained parameter values, which are very close to those
corresponding to this solution, as we mentioned in Sec.
\ref{kamioresults}. This could mean both that the two parameter sets
correspond to two distinct minima or to the same one. We will here follow
the latter assumption, {\it i.e.}, we will assume that there is one
minimum of the function (\ref{minfunction}) somewhere between these
two parameter regions. To avoid obtaining the mass squared differences
on the boundary between the two regions, we widened the interval for
$\Delta M^2$ to $10^{-4}\;{\rm eV}^2\leq \Delta M^2 \leq 10^{-2}\;{\rm
eV}^2$.  In the case of Solution 2, we restricted both mass parameters
to the same intervals as in Sec. \ref{kamiomini}, {\it i.e.}, to
$10^{-4}\;{\rm eV}^2\leq \Delta m^2 \leq 10^{-3}\;{\rm eV}^2$ and
$1\;{\rm eV}^2\leq \Delta M^2 \leq 10\;{\rm eV}^2$.

\begin{figure}[!ht]
\centering
\epsfig{figure=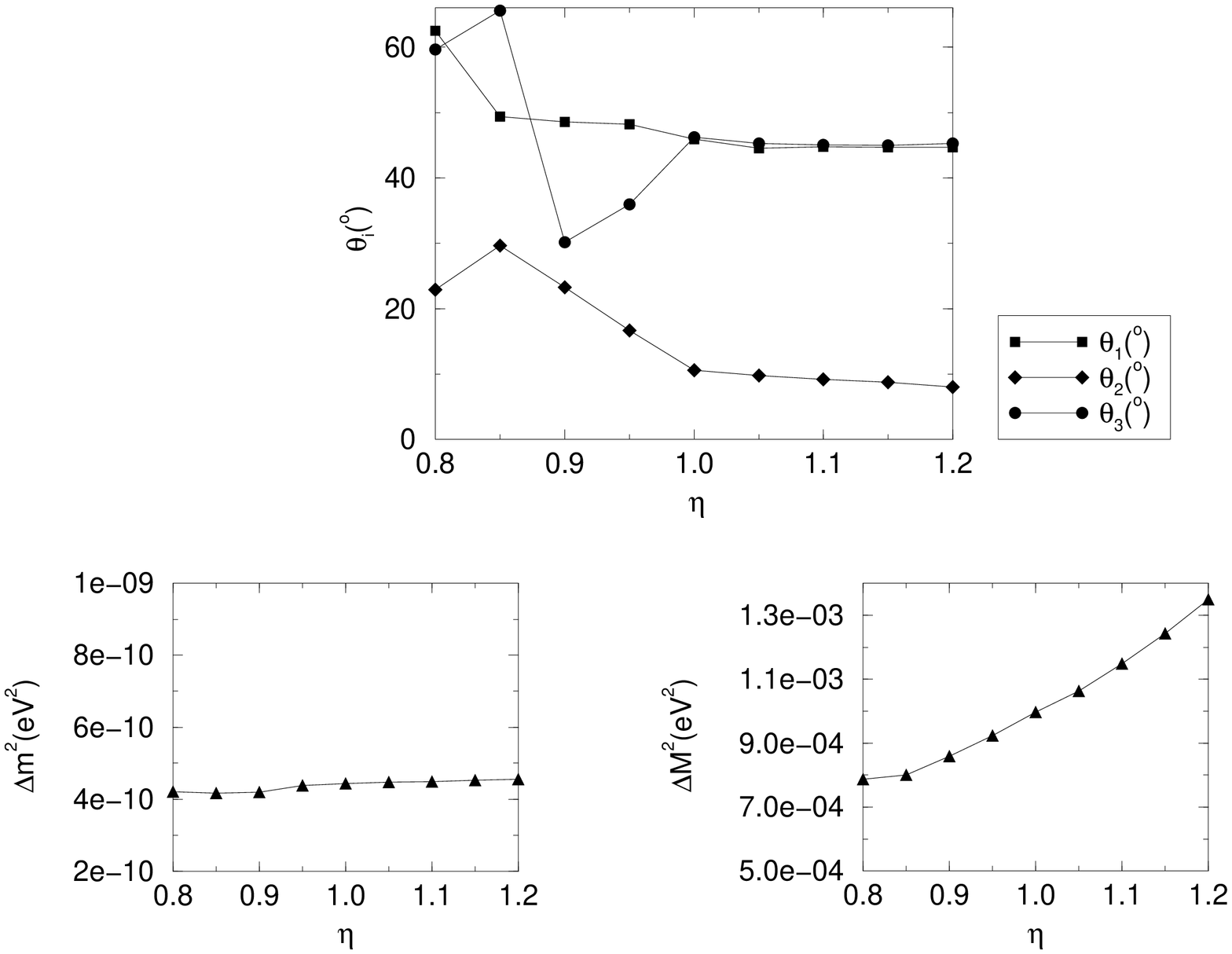,clip=,angle=-90,width=10cm}
\caption{Stability of Solution 1 with respect to variation of the flux
uncertainty factor $\eta$.}
\label{stabdiagram1}
\end{figure}

Figure \ref{stabdiagram1} shows the best point values for the five
parameters, which were obtained for the different values of the flux
uncertainty factor $\eta$ in the case of Solution 1. All three mixing angles
show the largest deviations from the value obtained for $\eta =1$ for
the smallest values of the flux uncertainty factor, whereas for
$\eta > 1$ they become almost stable. Somewhat surprising is the high
stability of the small mass squared difference, since it was obtained
from the minimization procedure with large standard deviations. The
large mass squared difference, finally, grows almost linearly with the
flux uncertainty factor, but remains within a region around $\Delta
M^2 = 10^{-3}\;{\rm eV}^2$.

\begin{figure}[!h]
\centering
\epsfig{figure=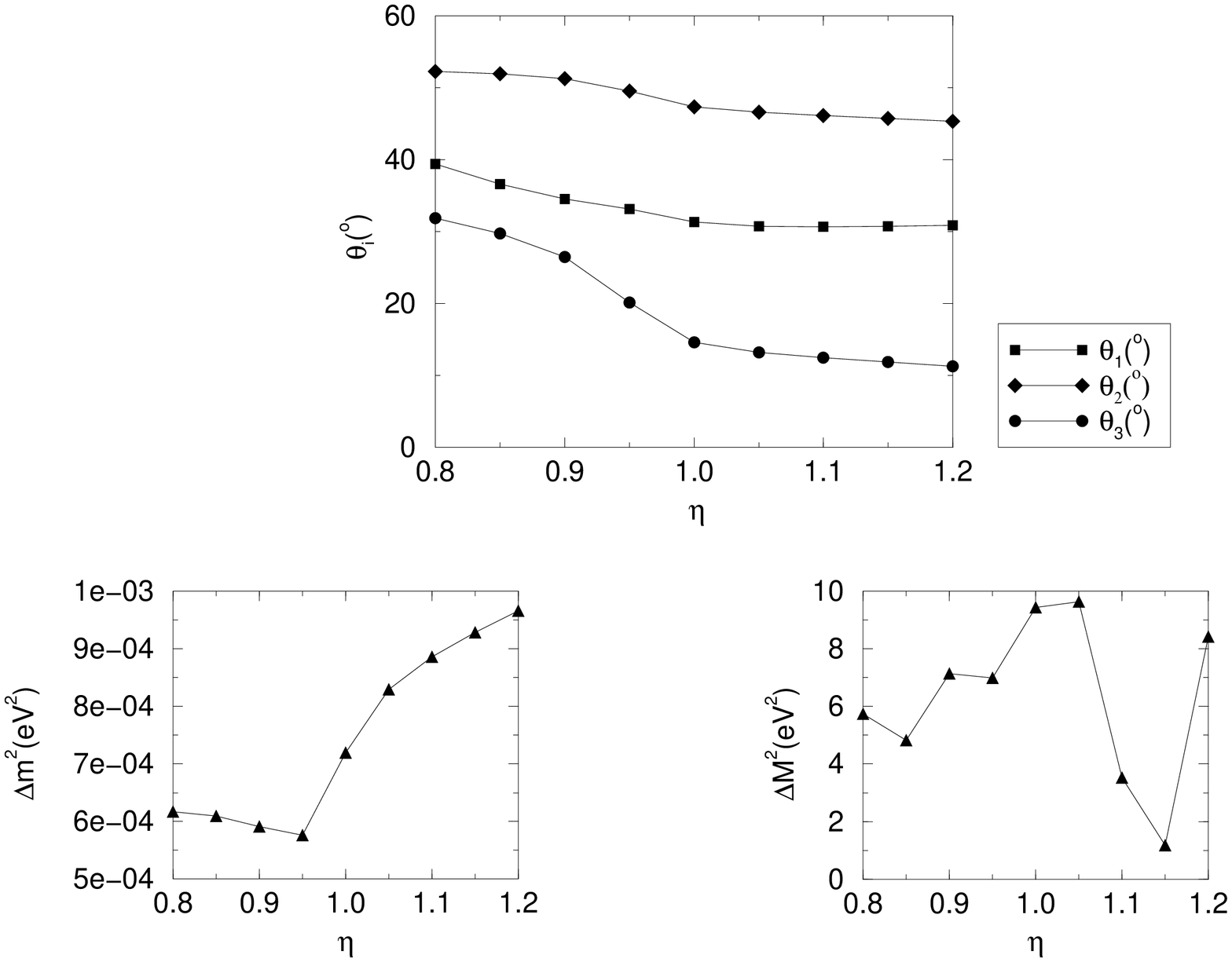,clip=,angle=-90,width=10cm}
\caption{Stability of Solution 2 with respect to variation of the flux
uncertainty factor $\eta$.}
\label{stabdiagram2}
\end{figure}

The dependence of the parameters
corresponding to Solution 2 on the flux uncertainty factor is shown
in Fig. \ref{stabdiagram2}. The mixing angles are clearly seen to be more
stable than in the case of Solution 1, they remain within ranges of
$10^\circ$ or less around their values at $\eta
=1$. The small mass squared difference, on the other hand, varies much
more than the one corresponding to the Solution 1. Finally, the
large mass squared difference shows a random distribution, as expected,
since the minimization function is independent of this parameter in
the mass region corresponding to Solution 2, as pointed out in Sec.
\ref{kamioresults}.

\section{Summary and Conclusions}
\label{kamioconclusions}

We have fitted the five parameters of a three flavor neutrino
oscillation scenario to experimental values obtained by the
Super-Kamiokande collaboration for atmospheric and solar neutrinos. To
obtain numerically stable solutions, we divided a large region for the
mass squared difference parameters into 45 smaller regions and
performed the fit for each of these regions. A mass region was
considered to contain a solution of the minimization problem, if the
best point values for mass squared differences were obtained from the
fit within the region boundaries. Furthermore, the best point value of
the minimization function was supposed to go through a minimum in such
a region if compared to the values obtained in the surrounding
regions. In this way, three types of solutions of the minimization
problem were obtained, out of which the best one corresponds to the
most common three flavor neutrino oscillation scenario, with one mass
parameter in the ``just-so'' range and the other one in the interval
$10^{-4}\;{\rm eV}^2\leq \Delta M^2 \leq 10^{-3}\;{\rm eV}^2$. As in
this common scenario, the first and the third mixing angle of this solution
were obtained to be maximal, $\theta_1 \simeq 45^\circ$ and $\theta_3
\simeq 45^\circ$, respectively, whereas the second one was obtained to
be small, $\theta_2 \simeq 10^\circ$, which is below the CHOOZ upper
bound. This solution corresponds to the
global minimum of the minimization function in the considered
parameter range, and it was the only one that provides a good fit to
all three types of data considered. As a disadvantage the value of the
large mass squared difference was obtained with a rather large standard
deviation.

The second solution is characterized by a value of
the small mass squared difference in the range, which contains
possible solutions of the atmospheric neutrino problem, {\it i.e.}, between 
$10^{-4}\;{\rm eV}^2$ and $10^{-3}\;{\rm eV}^2$ but by a
large mass squared difference in the LSND range, {\it i.e.}, between 1
${\rm eV}^2$ and 10 ${\rm eV}^2$. Here the mixing angles $\theta_1$
and $\theta_2$ were obtained to be maximal, whereas for the third one
there was obtained a smaller value, $\theta_3 \simeq 15^\circ$. This
solution provides a comparably worse fit to the atmospheric electron
neutrino data, which is the reason for the larger value of the
minimization function corresponding to that solution.

The third solution corresponds to a small mass squared difference in the range
$10^{-4}\;{\rm eV}^2\leq \Delta m^2 \leq 10^{-3}\;{\rm eV}^2$ and a
larger one in the LSND range. Here all three mixing angles were
obtained quite close to each other, $\theta_1\simeq
38^\circ,\;\theta_2 \simeq 29^\circ,\;\theta_3\simeq 25^\circ$, the
latter one with a rather large standard deviation. This solution
provides good fits to the atmospheric electron neutrino data and the
solar neutrino data. However, the oscillation probabilities for the
atmospheric muon neutrino data corresponding to this solution show no
zenith angle dependence such that this solution can be ruled out by the results
obtained by Super-Kamiokande.

\acknowledgments

Support for this work was provided by the Engineer Ernst Johnson
Foundation (T.O.). We would like to thank H{\aa}kan Snellman and Jonny
Lundell for useful discussions. Furthermore, we are grateful to Kate
Scholberg from the Super-Kamiokande Collaboration for providing us
with the experimental data.

\newpage

\begin{table}

\begin{tabular}{c|c|c|c|c|c}
Best point values for $\Delta m^2$ & $10^{-4} \leq \Delta M^2 \leq 10^{-3}$ & $10^{-3} \leq
\Delta M^2 \leq 10^{-2}$ & $10^{-2} \leq \Delta M^2 \leq 10^{-1}$ &
$10^{-1} \leq \Delta
M^2 \leq 1$ & $1 \leq \Delta M^2 \leq 10$\\
\hline
$10^{-11} \leq \Delta m^2 \leq 10^{-10}$ & $6.02 \times 10^{-11}$ &
$6.01 \times 10^{-11}$ & $5.94 \times 10^{-11}$ & $5.79 \times
10^{-11}$ & $5.77 \times 10^{-11}$\\
\hline
$10^{-10} \leq \Delta m^2 \leq 10^{-9}$ & $4.42 \times 10^{-10}$ & $4.44 \times
10^{-10}$ & $4.35 \times 10^{-10}$ & $4.25 \times 10^{-10}$ & $4.08
\times 10^{-10}$\\
\hline
$10^{-9} \leq \Delta m^2 \leq 10^{-8}$ & $10^{-9}$ & $10^{-9}$ & $10^{-8}$
& $10^{-9}$ & $10^{-8}$\\
\hline
$10^{-8} \leq \Delta m^2 \leq 10^{-7}$ & $10^{-8}$ & $1.04 \times
10^{-8}$ & $10^{-7}$ & $10^{-7}$ & $10^{-7}$\\
\hline
$10^{-7} \leq \Delta m^2 \leq 10^{-6}$ & $10^{-7}$ & $10^{-7}$ &
$10^{-6}$ & $10^{-6}$ & $10^{-6}$\\
\hline
$10^{-6} \leq \Delta m^2 \leq 10^{-5}$ & $10^{-6}$ & $10^{-6}$ &
$10^{-5}$ & $10^{-5}$ & $10^{-5}$\\
\hline
$10^{-5} \leq \Delta m^2 \leq 10^{-4}$ & $10^{-5}$ & $10^{-5}$ &
$10^{-4}$ & $10^{-4}$ & $10^{-4}$\\
\hline
$10^{-4} \leq \Delta m^2 \leq 10^{-3}$ & $9.28 \times 10^{-4}$ &
$10^{-4}$ & $8.70 \times 10^{-4}$& $7.77 \times 10^{-4}$& $7.20 \times
10^{-4}$\\
\hline
$10^{-3} \leq \Delta m^2 \leq 10^{-2}$ & $10^{-3}$ & $10^{-3}$ &
$10^{-3}$ & $10^{-3}$ & $10^{-3}$\\
\end{tabular}

\vspace{5mm}

\begin{tabular}{c|c|c|c|c|c}
Best point values for $\Delta M^2$ & $10^{-4} \leq \Delta M^2 \leq
10^{-3}$ & $10^{-3} \leq \Delta M^2 \leq 10^{-2}$ & $10^{-2} \leq
\Delta M^2 \leq 10^{-1}$ & $10^{-1} \leq \Delta M^2 \leq 1$ & $1 \leq
\Delta M^2 \leq 10$\\
\hline
$10^{-11} \leq \Delta m^2 \leq 10^{-10}$ & $10^{-3}$ &
$1.03 \times 10^{-3}$ & $4.25 \times 10^{-2}$ & $10^{-1}$ & $1.33$\\
\hline
$10^{-10} \leq \Delta m^2 \leq 10^{-9}$ & $9.92 \times 10^{-4}$ &
$1.01 \times 10^{-3}$ & $3.89 \times 10^{-2}$& $10^{-1}$ & $6.00$\\
\hline
$10^{-9} \leq \Delta m^2 \leq 10^{-8}$ & $9.89 \times 10^{-4}$ &
$10^{-3}$ & $3.84 \times 10^{-2}$ & $10^{-1}$ & $1.59$\\
\hline
$10^{-8} \leq \Delta m^2 \leq 10^{-7}$ & $9.89 \times 10^{-4}$ &
$10^{-3}$ & $3.83 \times 10^{-2}$ & $10^{-1}$ & $7.54$\\
\hline
$10^{-7} \leq \Delta m^2 \leq 10^{-6}$ & $9.89 \times 10^{-4}$ &
$10^{-3}$ & $3.84 \times 10^{-2}$ & $10^{-1}$ & $2.02$\\
\hline
$10^{-6} \leq \Delta m^2 \leq 10^{-5}$ & $9.89 \times 10^{-4}$ &
$10^{-3}$ & $3.84 \times 10^{-2}$ & $10^{-1}$ & $3.10$\\
\hline
$10^{-5} \leq \Delta m^2 \leq 10^{-4}$ & $9.82 \times 10^{-4}$ &
$10^{-3}$ & $3.82 \times 10^{-2}$ & $10^{-1}$ & $5.66$\\
\hline
$10^{-4} \leq \Delta m^2 \leq 10^{-3}$ & $10^{-4}$ & $10^{-3}$ &
$10^{-2}$ & $10^{-1}$ & $8.31$\\
\hline
$10^{-3} \leq \Delta m^2 \leq 10^{-2}$ & $10^{-4}$ & $1.30 \times
10^{-3}$ & $10^{-2}$ & $10^{-1}$ & $6.67$\\
\end{tabular}

\caption{Best point values of the mass parameters in the various
mass regions. All numerical values are given in units of ${\rm eV}^2$.}
\label{masstable}
\end{table}

\begin{table}
\begin{tabular}{c|c|c|c|c|c}
& $10^{-4} \leq \Delta M^2 \leq 10^{-3}$ & $10^{-3} \leq \Delta M^2
\leq 10^{-2}$ & $10^{-2} \leq \Delta M^2 \leq 10^{-1}$ & $10^{-1} \leq
\Delta M^2 \leq 1$ & $1 \leq \Delta M^2 \leq 10$\\
\hline
$10^{-11} \leq \Delta m^2 \leq 10^{-10}$ & $m$ & $M\;m$ & $M\;m$ & $m$
& $M\;m$\\
\hline
$10^{-10} \leq \Delta m^2 \leq 10^{-9}$ & $M\;m$ & $M\;m$ & $M\;m$ &
$m$ & $M\;m$\\
\hline
$10^{-9} \leq \Delta m^2 \leq 10^{-8}$ & $M$ & & $M$ & & $M$\\
\hline
$10^{-8} \leq \Delta m^2 \leq 10^{-7}$ & $M$ &$m$ & $M$ & & $M$\\
\hline
$10^{-7} \leq \Delta m^2 \leq 10^{-6}$ & $M$ & & $M$ & & $M$\\
\hline
$10^{-6} \leq \Delta m^2 \leq 10^{-5}$ & $M$ & & $M$ & & $M$\\
\hline
$10^{-5} \leq \Delta m^2 \leq 10^{-4}$ & $M$ & & $M$ & & $M$\\
\hline
$10^{-4} \leq \Delta m^2 \leq 10^{-3}$ & $m$ & & $m$ & $m$ &$M\;m$\\
\hline
$10^{-3} \leq \Delta m^2 \leq 10^{-2}$ & &$M$ & & & $M$\\
\end{tabular}

\caption{Lines of values for the mass squared differences obtained
within the boundaries. A capital ``$M$'' denotes a value for $\Delta
M^2$ obtained within the boundaries, whereas ``$m$'' stands for a
corresponding value for $\Delta m^2$. The regions labeled by
``$M\;m$'' contain solutions of the minimization problem. All
numerical values are given in units of ${\rm eV}^2$.}
\label{masstable2}
\end{table}

\begin{table}
\begin{tabular}{c|c|c|c|c|c}
Best point values for $\chi$ & $10^{-4} \leq \Delta M^2 \leq 10^{-3}$
& $10^{-3} \leq \Delta M^2 \leq 10^{-2}$ & $10^{-2} \leq \Delta M^2
\leq 10^{-1}$ & $10^{-1} \leq \Delta M^2 \leq 1$ & $1 \leq \Delta M^2
\leq 10$\\
\hline
$10^{-11} \leq \Delta m^2 \leq 10^{-10}$ &$7.2803 \times 10^{-3}$ &
$7.2842 \times 10^{-3}$ & $9.6916 \times 10^{-3}$ & $ 1.1829\times
10^{-2}$ & $1.9867 \times 10^{-2}$\\
\hline
$10^{-10} \leq \Delta m^2 \leq 10^{-9}$ & $3.7700 \times 10^{-3}$ &
$3.7677 \times 10^{-3}$ & $7.4108 \times 10^{-3}$ & $1.0046 \times
10^{-2}$ & $1.9245 \times 10^{-2}$\\
\hline
$10^{-9} \leq \Delta m^2 \leq 10^{-8}$ & $3.8319 \times 10^{-3}$ &
$3.8337 \times 10^{-3}$ & $7.4992 \times 10^{-3}$ & $1.0130 \times
10^{-2}$ & $1.9360 \times 10^{-2}$\\
\hline
$10^{-8} \leq \Delta m^2 \leq 10^{-7}$ & $3.8319 \times 10^{-3}$ &
$3.8338 \times 10^{-3}$ & $7.4992 \times 10^{-3}$ & $1.0970 \times
10^{-2}$ & $1.9360 \times 10^{-2}$\\
\hline
$10^{-7} \leq \Delta m^2 \leq 10^{-6}$ & $3.8320 \times 10^{-3}$ &
$3.8338 \times 10^{-3}$ & $7.4991 \times 10^{-3}$ & $1.0130 \times
10^{-2}$ & $1.9360 \times 10^{-2}$\\
\hline
$10^{-6} \leq \Delta m^2 \leq 10^{-5}$ & $3.8322 \times 10^{-3}$ &
$3.8342 \times 10^{-3}$ & $7.4971 \times 10^{-3}$ & $1.0127 \times
10^{-2}$ & $1.9353 \times 10^{-2}$\\
\hline
$10^{-5} \leq \Delta m^2 \leq 10^{-4}$ & $3.8345 \times 10^{-3}$ &
$3.8389 \times 10^{-3}$ & $7.3147 \times 10^{-3}$ & $9.8697 \times
10^{-3}$ & $1.8603 \times 10^{-2}$\\
\hline
$10^{-4} \leq \Delta m^2 \leq 10^{-3}$ & $3.8501 \times 10^{-3}$ &
$3.9166 \times 10^{-3}$ & $4.1743 \times 10^{-3}$ & $4.8230 \times
10^{-3}$ & $6.6859 \times 10^{-3}$\\
\hline
$10^{-3} \leq \Delta m^2 \leq 10^{-2}$ & $3.9166 \times 10^{-3}$ &
$4.0943 \times 10^{-3}$ & $4.2677 \times 10^{-3}$ & $5.1945 \times
10^{-3}$ & $7.2727 \times 10^{-3}$\\
\end{tabular}
\caption{Best point values of the minimization function in the various
mass regions. The mass squared differences $\Delta m^2$ and $\Delta
M^2$ are given in units of ${\rm eV}^2$.}
\label{minimizationtable}
\end{table}

\begin{table}
\begin{tabular}{cccccc}
\multicolumn{5}{l}{Solution 1}&\multicolumn{1}{r}{$\chi \approx 3.7677\times 10^{-3}$}\\
\hline
\hline
\makebox[1.5cm]{}&\makebox[1.5cm]{$\theta_1 (^\circ)$} &
\makebox[1.5cm]{$\theta_2 (^\circ)$} & \makebox[1.5 cm]{$\theta_3 (^\circ)$} & \makebox[1.5cm]{$\Delta m^2 ({\rm eV^2})$}
&\makebox[1.5cm]{$\Delta M^2 ({\rm eV^2})$}\\
\hline
Best point & 45.98 & 10.43 & 45.62 & $4.44 \times 10^{-10} $& $ 1.01
\times 10^{-3}$\\
\hline
Average with errors &$45.17 \pm 1.86$ & $11.03 \pm 0.85$ & $43.67 \pm 2.87$ &
$(6.64 \pm 2.13) \times 10^{-10}$ & $(1.02 \pm 0.02)\times 10^{-3}$\\
\end{tabular}

\vspace{5mm}

\begin{tabular}{cccccc}
\multicolumn{5}{l}{Solution 2} & \multicolumn{1}{r}{$\chi \approx
6.6859 \times 10^{-3}$}\\
\hline
\hline
\makebox[1.5cm]{}&\makebox[1.5cm]{$\theta_1 (^\circ)$} & \makebox[1.5cm]{$\theta_2(^\circ)$}
 & \makebox[1.5 cm]{$\theta_3 (^\circ)$} & \makebox[1.5cm]{$\Delta m^2 ({\rm eV^2})$}
&\makebox[1.5cm]{$\Delta M^2 ({\rm eV^2})$}\\
\hline
Best point & 31.34 & 47.38 & 14.64 & $7.20 \times 10^{-4}$&$8.13$\\
\hline
Average with errors & $31.35 \pm 0.02$ & $47.37 \pm 0.02$ & $14.62 \pm 0.05$ &
$(7.20 \pm 0.01) \times 10^{-4}$ & $6.30 \pm 2.48$ \\
\end{tabular}

\vspace{5mm}

\begin{tabular}{cccccc}
\multicolumn{5}{l}{Solution 3} & \multicolumn{1}{r}{$\chi \approx
1.9245 \times 10^{-2}$}\\
\hline
\hline
\makebox[1.5cm]{}&\makebox[1.5cm]{$\theta_1(^\circ)$} & \makebox[1.5cm]{$\theta_2(^\circ)$}
 & \makebox[1.5 cm]{$\theta_3(^\circ)$} & \makebox[1.5cm]{$\Delta m^2 ({\rm eV^2})$}
&\makebox[1.5cm]{$\Delta M^2 ({\rm eV^2})$} \\
\hline
Best point & 38.47 & 29.39 & 24.72 & $4.08\times 10^{-10}$ & 6.00\\ 
\hline
Average with errors & $38.22 \pm 0.49$ & $29.09 \pm 1.27$ & $34.50 \pm 17.61$ &
$(4.34 \pm 0.97) \times 10^{-10}$ & $6.03 \pm 2.65$\\
\end{tabular}

\caption{Parameter values for the three obtained solutions.}
\label{solutions}
\end{table}

\end{document}